\documentclass[letterpaper, 10 pt, conference]{IEEEtran}  % Comment this line out
\IEEEoverridecommandlockouts                              % This command is only
\usepackage{amsmath,amssymb,bm,bbm}

\usepackage{color, subfig}
\usepackage{graphicx}
\usepackage{tabularx}
\usepackage{epstopdf}
\usepackage{epsfig}
\usepackage{graphics} % for pdf, bitmapped graphics files
\usepackage{epstopdf}
\usepackage{enumerate}
\usepackage{caption}
\usepackage{eurosym}
\usepackage{amsmath} % assumes amsmath package installed
\usepackage{amssymb, dsfont}  % assumes amsmath package installed
\usepackage{subfig,pgfplots}
\pgfplotsset{ % Here we specify options for all figures in the document
  compat=newest, % Which version of pgfplots do we want to use?
  colormap={redblue}{ rgb255(0cm)=(128,0,0); rgb255(.4cm)=(255,0,0);rgb255(.8cm)=(255,255,0); rgb255(1.2cm)=(100,255,0);rgb255(3cm)=(0,255,255);rgb255(5cm)=(0,0,180)},
  legend style =
  {font=\small \sffamily},
  label style = {font=\small\sffamily},
every tick label/.append style={font=\small}}
\usepackage{tikz}

\usetikzlibrary {positioning}

\def\beq{\begin{equation}}
\def\eeq{\end{equation}}
\newtheorem{example}{Example}
\newtheorem{remark}{Remark}%}
\graphicspath{{images/}}

\newcommand{\R}{\mathbb{R}}

\newcommand{\N}{\mathbb{N}}

\newcommand{\mc}[1]{\mathcal{#1}}

\newcommand{\tup}[1]{\textup{#1}}
\newcommand{\bs}[1]{\boldsymbol{#1}}

\newcommand{\QED}{\hfill $\blacksquare$}

\usepackage{mathtools}
\usepackage[acronym]{glossaries}
\newacronym{SIS}{SIS}{Susceptible--Infected--Susceptible}
\newacronym{NPI}{NPIs}{nonpharmaceutical interventions}
\newacronym{MIP}{MIP}{Mixed-Integer Programming}
\newacronym{SoC}{SoC}{State of Charge}
\newacronym{PEV}{PEV}{Plug-in Electric Vehicle}
\newacronym{EV}{EV}{Electric Vehicle}
\newacronym{MLD}{MLD}{Mixed-Logical-Dynamical}
\newacronym{GS}{GS}{Gauss-Southwell}
\newacronym{GNEP}{GNEP}{Generalized Nash Equilibrium Problem}
\newacronym{MI-GPG}{MI-GPG}{Mixed-Integer Generalized Potential Game}
\newacronym{MINE}{$\varepsilon$-MINE}{$\varepsilon$-Mixed-Integer Nash Equilibrium}
\newacronym{CTM}{CTM}{Cell Transmission Model}
\newacronym{CTMs}{CTM-\textit{s}}{ Cell Transmission Model with service station}
\newacronym{CS}{CS}{Charging Station}
\newacronym{FV}{FV}{fuel vehicle}
\newacronym{r2s}{$\mathrm{r2s}$}{road-to-station}
\newacronym{s2r}{$\mathrm{s2r}$}{station-to-road}
\newacronym{HO}{HO}{Highway Operator}
\newacronym{TDM}{TDM}{Traffic Demand Management}
\newacronym{ATDM}{ATDM}{Active Traffic Demand Management}
\newacronym{NDW}{NDW}{Nationaal
Dataportaal Wegverkeer}
\newacronym{TTT}{TTT}{Total Travel Time}
\usepackage{bbold}
\usepackage{balance}

\title{\LARGE \bf
A Novel Control-Oriented Cell Transmission Model\\ Including Service Stations on Highways}

\author{Carlo Cenedese$^{\textrm{(a)}}$, Michele Cucuzzella$^{\textrm{(b)}}$, 
Antonella Ferrara$^{\textrm{(b)}}$, John Lygeros$^{\textrm{(a)}}$ %, Luca Vaghi$^{\textrm{(b)}}$% <-this % stops a space
\thanks{$^{\textrm{(a)}}$ Department of Information Technology and Electrical Engineering, ETH Z\"urich, Zurich, Switzerland ({\texttt{\{ccenedese,jlygeros\}@ethz.ch}}).
$^{\textrm{(b)}}$ 
Department of Electrical, Computer and Biomedical Engineering, University of Pavia, Pavia, Italy ({\texttt{\{michele.cucuzzella,antonella.ferrara\}@unipv.it}}).
The work of Cenedese  and Lygeros was supported by NCCR Automation, a National Centre of
Competence in Research, funded by the Swiss National Science
Foundation (grant number $180545$)}
}

\begin{document}

\maketitle

%%%%%%%%%%%%%%%%%%%%%%%%%%%%%%%%%%%%%%%%%%%%%%%%%%%%%%%%%%%%%%%%%%%%%%%%%%%%%%%%
\begin{abstract}
In this paper, we propose a novel model that describes how the  traffic evolution on a highway stretch is affected by the presence of a service station. The presented model enhances the classical \gls{CTM} dynamics by adding the dynamics associated with the service stations, where the vehicles may stop before merging back in the main stream. We name it \mbox{\gls{CTMs}}. We discuss its flexibility in describing different complex scenarios where multiple stations are characterized by different drivers' average stopping times corresponding to different services. The model has been developed to help designing control strategies aimed to decrease traffic congestion. Thus, we discuss how classical control schemes can interact with the proposed \gls{CTMs}. Finally, we validate the proposed model through numerical simulations and assess the effects of service stations on traffic evolution, which appear to be beneficial especially for relatively short congested periods.%  we use the data relating to a highway stretch between The Hague and Rotterdam, in The Netherlands to identify the model parameters.}
\end{abstract}

%%%%%%%%%%%%%%%%%%%%%%%%%%%%%%%%%%%%%%%%%%%%%%%%%%%%%%%%%%%%%%%%%%%%%%%%%%%%%%%%
\section{Introduction}\label{sec:intro}
In  recent years, mobility is becoming a central issue  in many countries and some alarming statistics show a growing need for change. For example, some studies show that the cost of congestion for the EU  is no less than $ 267$ billion {\euro } per  year~\cite{eu:2019:traffic}. Moreover, an inefficient transportation system affects not only the citizens' well-being, but also the environment, since traffic jams heavily increase the emission of $\text{CO}_2$ \cite{barth:2009:traffic}.
The classical solution to the traffic demand management problem is to increase  roads capacity, or to build alternative routes. Although  this solution produces tangible benefits~\cite{ganine:2017:resilience}, policymakers and researchers are exploring alternative interventions that may be  faster  and cheaper to implement.
These solutions heavily rely on  mathematical models to be able to  assess a priori their feasibility and impact. 

In traffic systems, the introduction of traffic models dates back to the 50s, with publication of the Lighthill-Whitham-Richards (LWR) model~\cite{lighthill:1955:LWR}, which is a  macroscopic model based on the equation of vehicles conservation representing traffic dynamics in an aggregate way. %Macroscopic traffic models are used to represent traffic dynamics in an aggregate way, exploiting the analogy between vehicles flow along road stretches and the flow of fluids or gases in pipelines. 
An alternative to macroscopic models is given by microscopic or mesoscopic traffic models~\cite[Sec.~2.3.1]{ferrara2018freeway}. The firsts explicitly capture drivers’ individual behaviors, while the seconds also account for car-following and vehicles interaction phenomena. Microscopic or mesoscopic traffic models are not convenient when the goal is to design traffic control strategies due to their computational complexity, definitely much higher than that of macroscopic models. % The latter, when adopted in a control design framework, are often formulated as discrete in time and space models. 
One of the most renowned  macroscopic discrete traffic models is the so-called \gls{CTM}, developed in the 90s for highway traffic in road stretches and networks \cite{daganzo:1995:CTM_part2}. Several variations of such a model has been developed throughout the years to address for its limitations~\cite{Kontorinaki:2017:CTM_capacity_drop,gomes:2006:asymmetric_CTM,kamonthep:2010:multiclass_CTM}. % It represents the traffic system as a collection of cells, some also able to describe merge and diverge traffic behaviors, where the traffic density evolution is modelled via a first-order finite difference equation.

The role of these models has been of paramount importance in the development of active traffic demand management mechanisms. Such  interventions may be designed to incentivize positive driver behaviors, and have their roots in behavioral economics and psychology~\cite{goodwin:2008:traffic_soft_measures}. Recently, inspired by the so-called ``valley filling'' objective in smart grids (see e.g. \cite{cenedese:2019:PEV_MIG,J_Hiskens_2013}) and  ramp metering control in highways, the authors of  \cite{cenedese:arxiv:highway_part_I,cenedese:arxiv:highway_part_II} have proposed an monetary incentive based policy for plug-in hybrid and electric  vehicles to alleviate traffic congestion. 
Another approach is to intervene by imposing some physical constraints or penalizing undesired actions. This category includes ramp metering, traffic lights and tolls control. The interested reader is referred  to~\cite{SIRI2021109655,depalma:2011:pricing} and the references therein for the details.

In this paper, we propose a novel highway traffic model that includes the dynamics of any service station along the highway. The presented model enhances the classical \gls{CTM} dynamics by modelling the presence of service stations where the vehicles may stop and then merge back in the main stream. The resulting  Cell Transmission Model with
service station (\gls{CTMs}) is a macroscopic highway traffic model capable of describing the dynamical effect of service stations on highway traffic. For example, the \gls{CTMs} can model a single service station that provides different services (refueling, ancillary services, charging for electric vehicles), or multiple service stations, that can be useful to study interventions similar to those in  \cite{ferguson:2020:carrots_or_sticks} where the authors propose an incentive-based approach to influence users' behavior in routing problems.
We discuss possible control schemes that  leverage  the \gls{CTMs} to perform active traffic demand management. For each  of these schemes, we discuss how it can be implemented and which would be the control actions put in place. Finally, we show via simulations that the presence of a service station can have a beneficial effect on highway traffic congestion, especially when short congestions occur.

% We believe that the  \gls{CTMs} can be exploited to design optimal model-based control schemes aimed at alleviating the traffic congestion and especially shaving the traffic peaks by influencing how the drivers interact with the service station. Among all possible control strategies, we think that soft incentive-based control policies might be very effective. For example, drivers can be encouraged to stop at service stations during congestion times by designing suitable discounts: the \gls{CTMs} can be used to identify the optimal discounts in terms of congestion alleviation. %for charging, food, drinks and newspapers.

\section{The \gls{CTM} with service stations }
\begin{figure}[t]
\centering
\includegraphics[trim=0 100 15 120,clip,width=\columnwidth]{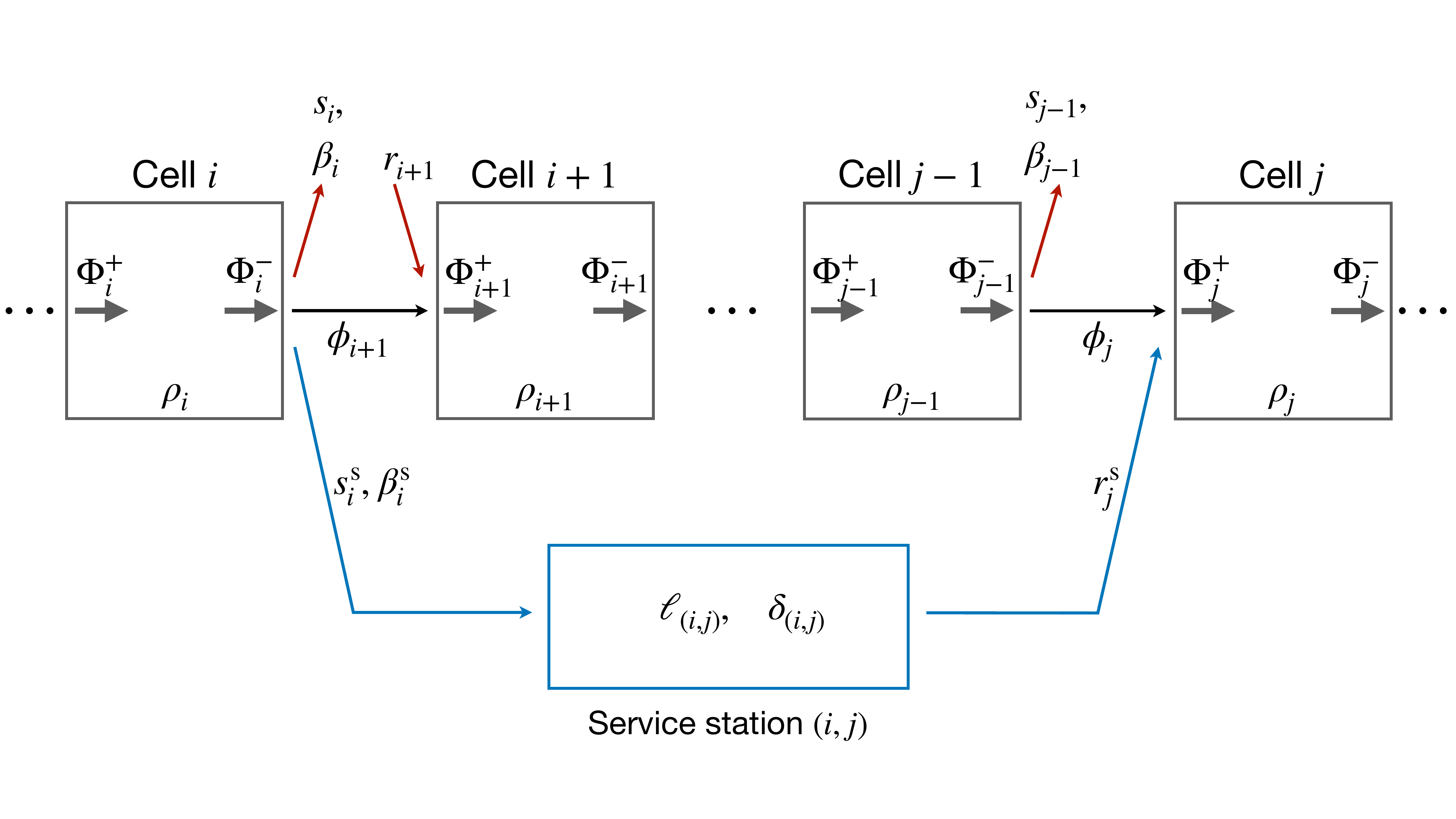}
\caption{The model and variables of the \gls{CTMs}. Here, there are on- and off-ramps together with a single service station $(i,j)$. }\label{fig:CTMs_scheme}
\end{figure}

 We model a generic highway stretch in which there may be on- and off-ramps that allow the vehicles to exit and merge into the main stream, respectively. Furthermore, we assume the presence of at least one  multi-purpose service station where a fraction of the drivers in the main stream stops to refuel or use   ancillary services such as the restaurant or restroom, see Figure~\ref{fig:CTMs_scheme}.  These vehicles obey different dynamics from those simply entering or exiting the highway: after a certain amount of time spent at the service station, they merge back in the main stream. This creates a coupling between the two flows that has to be carefully modelled. % to achieve a satisfactory description of the traffic evolution.     

% One of the most used models for   highway traffic control is the \gls{CTM} due to its simplicity and ability to capture the fundamental traffic dynamics~\cite{daganzo:1995:CTM_part2}. 
%The model owes its popularity to its simplicity and its ability to describe  macroscopic traffic dynamics for simple road networks, see~\cite{rinaldi:ferrara:2012:CTM_identification}. 
In the remainder of the section, we use the ``classical'' formulation of the \gls{CTM} as cornerstone to build the proposed  \gls{CTMs}. 

\subsection{Model variables}
 
\begin{figure}[t]
\centering
\includegraphics[trim=150 0 200 0,clip,scale=0.5,width=\columnwidth]{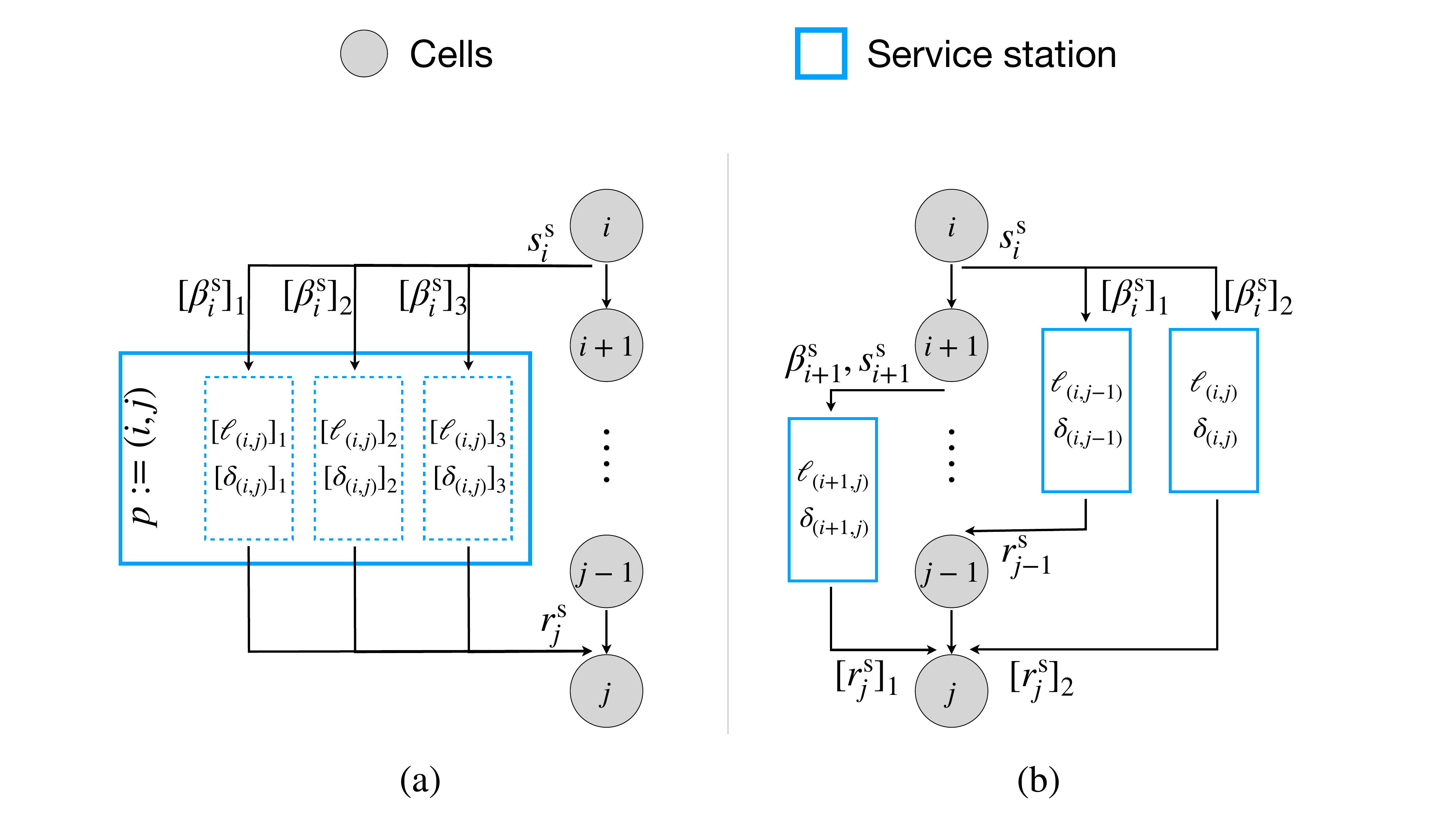}
\caption{Different configurations of the \gls{CTMs}. (a) A multi-modal service station where different $\delta_{(i,j)}$ represent different behaviours of the drivers. (b) A complex highway stretch where $(i,j-1)$ and $(i,j)$ share the entry point while $(i+1,j)$ and $(i,j)$ share the exit point.}\label{fig:CTMs_conf}
\end{figure}

%%%%%%%
Each time interval $[kT,(k+1)T)$ of length $T$ is denoted by an integer $k\in\N$.
 The highway stretch is modeled as a chain of $N$ consecutive \textit{cells} and the vehicles in each cell $i\in\mc N\coloneqq \{1,\dots,N\}$ are assumed to  move  with constant speed $v_i(k)$. Two adjacent cells are connected via an \textit{interface}, where the vehicles can \textit{i}) proceed to the next cell; \textit{ii}) exit the main stream via an off-ramp or by stopping at a service station; \textit{iii})~merge into the main stream of the next cell via an on-ramp or by exiting a service station. We assume the presence of $M$ service stations. Each station $p\in\mc M\subseteq \mc N\times \mc N$   is located  between any two cells $i,j\in\mc N$, and $|\mc M|=M$.  Station $p$ correspond to the ordered couple $(i,j)\in\mc M$, where  $i\in\mc N$ represents the cell from which the vehicles may access the service station, while $j\in\mc N$ denotes the one in which they merge back. Notice that multiple stations can be represented by the same couple, since  stations can  share enter and exit points, see Figure~\ref{fig:CTMs_conf}(a).
 %Let us introduce some specific notation used in the following.
 We denote the set of all the service stations having access  point at $i$  by \mbox{$\mc E^{\tup{in}}_i=\{p\in\mc M \,|\, \exists \, j\in\mc N \tup{ s.t. }  p=(i,j)  \}$.} Those that merge back into cell $i$ by $\mc E^{\tup{out}}_i$  and note that $\sum_{i\in\mc N} |\mc E^{\tup{in}}_i| = \sum_{i\in\mc N} |\mc E^{\tup{out}}_i| = M $.
 %Notice that the model developed does not assume that $i\neq j$, or that the service stations do not share the entering or the exiting cells. In the following, we discuss this powerful feature of the proposed model that allows to easily model various scenarios.   

Following  \cite[Sec.~3.3]{ferrara2018freeway}, we briefly introduce the variables used in the proposed \gls{CTMs} for a  generic cell $i\in\mc N$ and service station $p=(i,j) $, during an  interval $k\in\N$.  A set of fixed parameter is associated to each cell $i$:
\begin{itemize}
\item $L_i\:[\tup{km}]$: cell length;
\item $\overline v_i\:[\tup{km/h}]$: free-flow speed;
\item $w_i\:[\tup{km/h}]$: congestion wave speed;
\item $q_i^{\max}\:[\tup{veh/h}]$: maximum cell capacity;
\item $\rho_i^{\max}\:[\tup{veh/km}]$: maximum jam density;
% \item $r^{\max}_{i}\:[\tup{veh/h}]$: maximum capacity of the on-ramp sections;
\item $r^{\tup s,\max}_{i}\:[\tup{veh/h}]$: maximum capacity of the on-ramp exiting a service station.
\end{itemize}
The variables  used to describe the dynamics (during the time interval $k\in\N$) are:
\begin{itemize}
\item $	\rho_i(k)\: [\textup{veh/km}]$: traffic density of  cell $i$;
\item $\Phi_i^+(k) \:[\textup{veh/h}]$ (resp. $\Phi_i^-(k)$):  total flow entering (exiting)  cell $i$;
\item $\phi_i(k)\:[\textup{veh/h}]$: flow entering cell $i$ from  $i-1$; %$\phi_{1}(k)$ (resp. $\phi_{N+1}(k)$) is the flow entering (exiting) the highway during the same interval;
\item $r_i(k)\:[\textup{veh/h}]$: flow of vehicles merging into the main stream via an on-ramp;
\item $r^{\tup s}_i(k)\:[\textup{veh/h}]$:  flow of vehicles merging into the main stream from the service stations in $\mc E^\tup{out}_i$;
\item $s_i(k)\:[\textup{veh/h}]$:  flow of vehicles leaving the main stream via an off-ramp;
\item $s^{\tup s}_i(k)\:[\textup{veh/h}]$: flow of vehicles leaving the main stream to enter  the service stations in $\mc E^\tup{in}_i$;
\item $\beta_i(k)$:  split ratio associated with the off-ramp;
\item $\beta^{\tup s}_i(k)$:  split ratio associated with the service stations in $\mc E^\tup{in}_i$;
\item $\ell_{(i,j)}(k)$ (or  $\ell_p$) $\:[\textup{veh}]$: number of vehicles at the service station $(i,j)$;
\item $e_{(i,j)}(k)$ (or $e_p$) $\:[\textup{veh}]$:  number of vehicles queuing at the service station $(i,j)$ due to the impossibility of merging back in the main stream during previous time intervals;
%\item $r^{\tup {s,c}}_{i,j}(k)\:[\textup{veh/h}]$: flow of vehicles attempting to exit the service station $(i,j)$ during $k$;
\item $\delta_{(i,j)}$ (or  $\delta_p$): the average number of time intervals spent at the service station $(i,j)$ by the drivers. It is assumed constant for the sake of simplicity, but in general may be time varying and we discuss this aspect in Section~\ref{sec:control};
\item $p_i$: relative  priority of all the flows entering in cell $i$ in case of congestion. By definition, this vector has to belong to a simplex, i.e., $\bs 1^\top p_i=1$ where $\bs 1$ is a vector of all $1$ of the correct dimension. The  component of $p_i$ associated with the main stream is denoted by $p^\tup{ms}_i>0$. For simplicity, we assume this vector constant over time and discuss this in Section~\ref{sec:control}. 
\end{itemize} 

In the remainder, given a variable $x$, we denote its components associated with $q\in\mc M$  by $[x]_q$. 

In general,  at the interface between cells $i$ and $i+1$  there can be the access and exit of multiple service stations.  Consequently, $\beta^{\tup s}_i\in\R^{|\mc E^{\tup{in}}_i|}$, where $[\beta^{\tup s}_i]_q$ represents the split ratio of vehicles entering the $q$-th station in $\mc E^{\tup{in}}_i$. Following a similar reasoning, we attain that  $r^{\tup s,\max}_{i},r^{\tup s}_i\in\R^{|\mc E^{\tup{out}}_i|}$ and $\ell_{(i,j)},\delta_{(i,j)}, e_{(i,j)}\in \R^{|\mc E^{\tup{in}}_i \cap \mc E^{\tup{out}}_j|} $.  
Finally, to ease the notation, we assume  that for a cell $j$ there cannot be in-flows deriving from both an on-ramp and the exiting of a service station. The role of this assumption will appear clear in Section~\ref{sec:CTMs_dyn}.

To clarify the role of the variables defined above, we present two possible configurations schematized in Figure~\ref{fig:CTMs_conf}. 
\smallskip
\begin{example}[Single multi-modal service station]
In Figure~\ref{fig:CTMs_conf}(a), we represent the case of  single service station where  drivers can stop for different reasons, e.g., refueling, ancillary services, or charging a plug-in electric vehicles. These different types of services are described via $M=3$ service stations that share the same entry and exit points, i.e., cells $i$ and $j$, respectively. So, $|\mc E^{\tup{in}}_i|=|\mc E^{\tup{out}}_j|=|\mc E^{\tup{in}}_i\cap \mc E^{\tup{out}}_j|=3$. This scheme allows us to use different split ratios $[\beta_i^\tup{s}]_q, q\in \mc M$, for the three different types of services, and  different $[\delta_{(i,j)}^{\tup s}]_q, q\in \mc M$, describing the average time spent at the corresponding service station.
\end{example}

\smallskip
\begin{example}[Multiple service stations]
In Figure~\ref{fig:CTMs_conf}(b), we represent a generic configuration in which there are $M=3$ service stations, i.e., $\mc M=\{m_1, m_2, m_3\}$, where $m_1=(i,j-1),\,m_2=(i,j),\,m_3=(i+1,j)$. Here, $m_1$, $m_2$ share the same entry point, so $\beta_i^{\tup s}\in\R^2$ since $\mc E^{\tup{in}}_i = \{ m_1,m_2 \}$, while $\beta_{i+1}^{\tup s}\in\R$. Similarly, $m_2$ and $m_3$ merge back in the same cell $j$, so $\mc E^{\tup{out}}_j = \{ m_2,m_3 \}$ and thus $r^{\tup s}_j\in \R^2$. Conversely, $|\mathcal{E}^{\tup{in}}_i \cap \mathcal{E}^{\tup{out}}_{j-1}|=|\mathcal{E}^{\tup{in}}_i \cap \mathcal{E}^{\tup{out}}_j|=|\mathcal{E}^{\tup{in}}_{i+1} \cap \mathcal{E}^{\tup{out}}_j|=1$ so $\ell_p, \,\delta_p \in \R$ for all $p\in\mc M$.
\end{example}

\subsection{\gls{CTMs} dynamics}\label{sec:CTMs_dyn}
We are now ready to introduce the dynamics of the proposed  \gls{CTMs}.  %that describe the evolution of the aforementioned variables.
 The evolution of the density $\rho_i$ of cell $i\in\mc N$  is governed by
\smallskip
\begin{equation}
\label{eq:rho_dyn}
\rho_i(k+1) = \rho_i(k) + \dfrac{T}{L_i}\left( \Phi_i^+(k)-\Phi_i^-(k)\right) \:,
\end{equation}
where the inflow and outflow are defined respectively as
\smallskip
\begin{subequations}
\label{eq:Phi_+_-}
\begin{align}
\label{eq:Phi_-}
\Phi^-_i(k) &\coloneqq 
 \phi_{i+1}(k) + s_i(k) + s^{\tup s}_i(k)\\ 
\Phi^+_i(k) &\coloneqq \phi_i(k) + r_i(k) + \boldsymbol 1^\top  r^{\tup s}_i(k), 
\end{align} 
\end{subequations}
and $\bs 1\in\R^{|\mc E^{\tup{out}}_i|}$. 

As in \cite[Sec. 3.3.1]{ferrara2018freeway}, the flow $s_i(k)$ is  a fraction $\beta_i$ of the total flow exiting the cell, so $s_i(k)=\frac{\beta_i(k)}{1-\beta_i(k)}\phi_{i+1}(k)$. 
Likewise, the total flow entering the service stations is defined as
\begin{equation}
    s^{\tup s}_{i}(k)=\sum_{q\in\mc E^{\tup{in}}_i}\frac{[\beta^{\tup s}_{i}(k)]_q}{1-[\beta^{\tup s}_{i}(k)]_q}\phi_{i+1}(k)=\sum_{q\in\mc E^{\tup{in}}_i}[\beta^{\tup s}_{i}(k)]_q\Phi^-_{i}(k) .
    \label{eq:s_c_i}
\end{equation}
The only constraint that has to be satisfied by the split ratios is $0\leq \beta_i(k)+\bs 1^\top \beta_i^{\tup s}(k) \leq 1$. Notice that we do not explicitly model the supply of the service station since, if needed, can be incorporated by imposing dynamic constraints on $\beta_i^\tup{s}(k)$.

The vehicles entering a service station $q=(i,j)$ during $k$ linger there for $\delta_{q}$ time intervals before trying to merge back. The number of vehicles at the service station evolves as 
\begin{align}
    \ell_{q}(k+1)=\ell_{q}(k)+T\left[ s_{q}^{\tup s}(k) -[r^{\tup s}_{j}(k)]_q\right],     \label{eq:l_dyn} 
\end{align}
 where  $s_{q}^{\tup s}(k)\coloneqq [\beta^{\tup s}_{i}(k)]_q \Phi_i^-(k)$ and $q\in \mc E^{\tup{in}}_i \cap \mc E^{\tup{out}}_{j}$. Loosely speaking, $s_{(i,j)}^{\tup s}(k)\in\R^{|\mc E^\tup{in}_i\cap\mc E^\tup{out}_j|}$ comprises  the fractions of $s^\tup{s}_{i}(k)$ in \eqref{eq:s_c_i} entering the service station $q$. 
 
After $\delta_q$ time intervals, the vehicles attempt to exit the service station to merge back into   the main stream. However, if this flow exceeds the supply of the receiving cell, then some vehicles remain at the service station and wait for merging back during the next intervals, thus creating a queue. To keep track of these vehicles, we introduce the state variable  $e_q(k)$, whose dynamics evolve as follows
 \begin{equation}\label{eq:e_q}
     e_q(k+1) = e_q(k) + T[s^{\tup s}_{q}(k-\delta_{q}) - r^{\tup s}_{j}(k)]_q\,.
 \end{equation}
 With a slight abuse of notation, we can compactly define the flow of vehicles that attempts to exit during $k$ as $s^{\tup s}_{q}(k-\delta_{q})+\frac{e_q(k)}{T}$.%, where the average  time $\delta_{q}$ spent at the service station can be interpreted as a delay.
 If the flow $s^{\tup s}_{q}(k-\delta_{q})+\frac{e_q(k)}{T}$ exceeds the capacity of the on-ramp connected to cell $j$, then only part of it is able to  merge back. Specifically, the demand of the ramp exiting the service station $(i,j)$ and connecting to cell  $j$ reads as
\begin{equation}
\label{eq:demad_ij}
    D^{\tup{s}}_{q}(k) = \min\left(s^{\tup s}_{q}(k-\delta_{q})+\frac{e_q(k)}{T},r_{q}^{\tup s,\max} \right) .
\end{equation}
\smallskip
\begin{remark}[Number of variables]
A drawback of the current formulation of the proposed \gls{CTMs} is the necessity of $\delta_{q}$  variables to track the evolution of $\ell_{q}$ over time. This might be overcome by rephrasing the dynamics of the service station via a stochastic variable that obeys a Poisson distribution. We leave the investigation of this alternative formulation to future works.   
\end{remark}
\smallskip
We denote the demand of the on-ramps of cell $j$ during $k$ by $D_j^\tup{ramp}(k)$. 
On the other hand, similarly to the classical \gls{CTM}, the demand of cell $i$ and the supply of cell $j$ are respectively
\begin{subequations}
\begin{align}
\label{eq:D_i}
    D_{i}(k) &= \min\left(\bigl(1-\beta_i(k)-\bs 1^\top \beta^{\tup s}_{i}(k)\bigr)\overline v_{i}\rho_{i}(k), q_{i}^{\max}\right),\\
\label{eq:S_j}
    S_{j}(k) &=\min\left(w_{j}(k)\bigl(\rho_{j}^{\max}(k)-\rho_{j}(k)\bigr),q_{j}^{\max}\right) .
\end{align}
\end{subequations}
We define next the relation between the demand/supply of cells and ramps and the flows of vehicles that transit from one to another.
Let us discuss individually the free-flow and the congested case for cell $j$. We only focus on the case in which there is an in-flow deriving from the exit of a service station, i.e., $|\mc E^{\tup{out}}_j|>0$, and refer the reader to \cite[Eq.~3.33-3.34]{ferrara2018freeway} for the case in which the flow is due to an on-ramp.
\subsubsection{Free-flow case} This is the simplest scenario and arises if $$D_{j-1}(k)+ \sum_{p\in \mc E^{\tup{out}}_j}D_{p}^\textup{s}(k)\le S_j(k).$$ 
Then, all the vehicles are able to enter cell $j$ during $k$, and thus the flows read as
\begin{subequations}
\begin{align}
    \phi_j(k) &= D_{j-1}(k)\\
    [r^\tup{s}_j]_{q} &= D_{q}^\tup{s} (k),\: \forall q\in\mc E_j^\tup{out} 
\end{align}
\end{subequations}

\subsubsection{Congested case}
The total demand exceeds the supply of cell $j$, i.e., $D_{j-1}(k)+ \sum_{p\in \mc E^{\tup{out}}_j}D_{p}^\textup{s}(k)> S_j(k)$. This might occur for different reasons, so we discuss them separately. 
\begin{itemize}
    \item
If $D_{j-1}(k)>p_j^\tup{ms}S_{j}(k)$ and
$\sum_{p\in \mc E^{\tup{out}}_j}D_{p}^\textup{s}(k)\leq (1-p_j^\tup{ms})S_{j}(k)$, then the complete flow from the stations can enter the cell $j$, so the flows read as
\begin{subequations}
\begin{align}    
\phi_j(k) &= S_j(k)-\sum_{p\in \mc E^{\tup{out}}_j}D_{p}^\textup{s}(k)\\
    [r^\tup{s}_j]_{q} &= D_{q}^\tup{s} (k),\: \forall q\in\mc E_j^\tup{out} .
\end{align}
\end{subequations}
\item If $D_{j-1}(k)\leq p_j^\tup{ms}S_{j}(k)$ and
$\sum_{p\in \mc E^{\tup{out}}_j}D_{p}^\textup{s}(k)> (1-p_j^\tup{ms})S_{j}(k)$, then the whole flow from the previous cell can enter cell $j$, thus
\begin{equation}
    \phi_j(k) =D_{j-1}(k).  
\end{equation}
The remainder of the supply is split  among the ramps exiting the service station, i.e., $\sum_{p\in \mc E^{\tup{out}}_j}D_{p}^\textup{s}(k) =S_j(k)-D_{j-1}(k)$. Finding the exact expression for the flow of each ramp must be done with an iterative procedure. First, we split the ramps of the stations $\mc E^\mathrm{out}_j$ in two subsets that are $\,^1\overline{\mc E}$ and $\,^1\underline{\mc E}$ ($1$ denotes the procedure's iteration). The former is composed by those ramps $p\in\mc E^\tup{out}_j$ satisfying 
$$D_q^\tup{s}>\frac{S_j(k)-D_{j-1}(k)}{|\mc E^\tup{out}_j|},$$
 while $\,^1\underline{\mc E}= \mc E^\tup{out}_j \setminus \,^1\overline{\mc E}$. From this, we conclude that the complete flow of the ramps in $\,^1\underline{\mc E}$ manages to  enter cell $j$ during $k$, thus
 \begin{equation}
   [r^\tup{s}_j]_{q} = D_{q}^\tup{s} (k),\: \forall p\in \,^1\underline{\mc E}.   
 \end{equation}
 Next, we define $\,^2\overline{\mc E}$ as those $p\in\,^2\overline{\mc E}$ that satisfy 
 $$D_q^\tup{s}>\frac{S_j(k)-D_{j-1}(k)-\sum_{p\in\,^1\underline{\mc E}}D_p^{\tup s}}{|\,^1\overline{\mc E}|},$$
 and $\,^2\underline{\mc E} = \,^1\overline{\mc E}\setminus \,^2\overline{\mc E}$.  Also in this case we obtain that  
  \begin{equation}
   [r^\tup{s}_j]_{q} = D_{q}^\tup{s} (k),\: \forall q\in\,^2\underline{\mc E}.  
 \end{equation}
By repeating this procedure, we  reach an iteration $y$ such that $\,^y\underline{\mc E}=\emptyset$. Therefore, the flow from the ramps in  $\,^y\overline{\mc E}$ cannot enter completely in cell $j$ and thus the remaining supply  
$$\hat{S}_j(k)\coloneqq S_j(k)-D_{j-1}(k)-\sum_{t=1}^{y}\sum_{p\in\,^t\underline{\mc E}}D_p^{\tup s}(k)$$
is divided among them based on their priority
\begin{equation}
\label{eq:r_s_end}
   [r^\tup{s}_j]_{q} =\dfrac{[p_j]_q }{\sum_{p\in\,^y\overline{\mc E}} [p_j]_p } \hat{S}_j(k),
 \end{equation}
for all $q\in \,^y\overline{\mc E}$.

\item If $D_{j-1}(k) > p_j^\tup{ms}S_{j}(k)$ and
$\sum_{q\in \mc E^{\tup{out}}_j}D_{q}^\textup{s}(k)> (1-p_j^\tup{ms})S_{j}(k)$, then
\begin{equation}
    \phi_j(k) =p_j^\tup{ms}S_{j}(k).  
\end{equation}
The total flow entering from the service stations in cell $j$ is $\sum_{p\in \mc E^{\tup{out}}_j}D_{p}^\textup{s}(k) =(1-p_j^\tup{ms})S_j(k)$. To obtain the single flows, we apply the same iterative procedure described above where we have to replace $S_j(k)-D_{j-1}(k)$ with $(1-p_j^\tup{ms})S_j(k)$ both in the sets definition and in \eqref{eq:r_s_end}. Therefore, the resulting flows read as 
\begin{align*}
    [r^\tup{s}_j]_{q} &= D_{q}^\tup{s} (k),\: \forall p\in\bigcup_{t=1}^y\,^t\underline{\mc E},\\
    \tilde S_j(k)&\coloneqq  (1-p_j)S_j(k)-\sum_{t=1}^{y}\sum_{q\in\,^t\underline{\mc E}}D_q^{\tup s} \\
   [r^\tup{s}_j]_{q} & =\dfrac{[p_j]_q }{\sum_{p\in\,^y\overline{\mc E}} [p_j]_p }\tilde S_j(k),\: \forall q\in\,^y\overline{\mc E}.  
\end{align*}
\end{itemize}

This concludes the formulation of the \gls{CTMs}, in fact all the dynamics associated with the variables introduced in the previous section have been defined. The formulation above, even though it might seem convoluted, boils down to very simple and intuitive equations when it is applied to specific cases like the ones depicted in Figure~\ref{fig:CTMs_conf}. 
% Due to space limitation, we omit
% simulations produced by implementing the proposed model in Aimsun Next (see Figure~\ref{fig:aimsun}). Their results will be discussed during the presentation of the work.

%Due to space limitation we omit simulations of the model that have been performed in Aimsun Next (see Fig. \ref{fig:aimsun}) and will be discussed during the presentation of the work.

% \begin{figure}[t]
% \centering
% \includegraphics[width=\columnwidth]{figure/aimsun.eps}
% \caption{The 2D and 3D microscopic views of freeway traffic simulation in Aimsun Next.}\label{fig:aimsun}
% \end{figure}

\smallskip
\begin{remark}[Capacity drops]
Historically the \gls{CTM} has some limitations. Arguably, the most important is the inability to model  capacity drops on the highway. To solve this problem, alternative formulations of the \gls{CTM} can be adopted, see e.g.
\cite{Kontorinaki:2017:CTM_capacity_drop,Han:2017:CTM_resolving_freeway_jam,Piacentini:2019:VACS_congestion_dissipation}. The  modifications developed for the proposed \gls{CTMs} can also be introduced into these more advanced versions of the \gls{CTM}. In fact, the key aspect of the \gls{CTMs} is the connection of on- and off-ramps of the service stations via queue dynamics, which does not affect the core dynamics of the model. 
\end{remark}

\section{Overview of possible control strategies }
\label{sec:control}
This section is devoted to  discuss  how the proposed \gls{CTMs} can be utilized for designing novel traffic control schemes. We provide a high-level description of   possible control techniques based on the \gls{CTMs} and identify the models' variables  that would be influenced by the control actions.

\subsubsection{Control of the flow exiting the service station}
the dynamics of the service station are inspired by those associated with on- and off-ramps. A natural idea is to introduce  a ramp-metering mechanism to modulate the flow of vehicles merging back into the main stream, i.e., $r^\tup{s}$. This can be implemented introducing a control signal $r^\tup{s,c}_q(k)$ in \eqref{eq:demad_ij}  for every $q\in\mc M$. Then, similarly to \cite[Eq.~3.39]{ferrara2018freeway}, the demand becomes 
\begin{equation*}
    D^{\tup{s}}_{q}(k) = \min\left(\frac{s^{\tup s}_{q}(k-\delta_{q})+e_q(k)}{T}, r^\tup{s,c}_q(k) ,r_{q}^{\tup s,\max} \right).
\end{equation*}
The design of $r^\tup{s,c}_q(k)$ can be performed  applying  ramp-metering control methods such as ALINEA \cite{papageorgiou:1997:ALINEA}, where the local control influences the exit flow based solely on the traffic conditions of the highway. Alternatively, one can take direct advantage of the knowledge of the \gls{CTMs} dynamics by introducing an MPC-based or event-triggered control scheme, such as those discussed in \cite[Sec.~5.3 and 5.4]{SIRI2021109655}.

The intuition behind these schemes is that  limiting the flow of vehicles merging back in the main stream during rush hours decreases the overall traffic congestion. Overall, these controls might improve the overall traffic congestion but could also  increase $e_q(k)$.  
  
\subsubsection{ Incentive based traffic control}
while the flow of vehicles exiting the service station can be directly controlled, e.g., via a toll, the control over the entering one cannot be performed in a direct way. In fact, the drivers passing by the entrance of the service stations can decide whether to stop or not. Thus, the value of $\beta(k)$ arises from the individuals' decisions. 

Game theory can be a suitable tool to analyze such phenomena since there is a rich literature that studies how a game can model and influence  decision-making processes~\cite{cenedese:2019:RBR,cenedese:2021:EJC}. In the game, the payoff would be associated with the current and future traffic conditions estimated via the \gls{CTMs}. Then, the decision can be influenced via incentives. The game's outcome during $k$ would define the value of  $\beta(k)$. A similar approach can also be used to influence $\delta(k)$. 
The nature of these incentives may be diverse and varies from a discounted energy price for the charging of electric vehicles \cite{cenedese:arxiv:highway_part_I} to several benefits for using the ancillary services. The effectiveness of using rewards to reinforce a desirable behavior is supported by a large volume of empirical evidence \cite{berridge:2000:reward,kreps:1997:extrinsic_incentives} and the proposed \gls{CTMs} can prove to be an important tool to shift from static incentives to dynamic  ones. 

\subsubsection{Optimal service station  positioning} the proposed  \gls{CTMs} has high flexibility and can easily describe many service stations configurations, by manipulating $\beta$, $\delta$ and $r^\tup{s}$. This  feature can be exploited to create computationally tractable optimization problems to encompass  the optimal configuration and positioning of service stations along a long and complex highway. This would improve  current studies that rely directly on micro-simulators that require the use of algorithms that cannot guarantee the solutions' optimality, e.g., the genetic algorithm applied to SUMO \cite{hess:2012:optimal_CS_pos}.

\medskip    

The above discussion is meant to highlight the many possible applications in which the use of the proposed \gls{CTMs} can be beneficial to design and predict effective and novel traffic control actions. Moreover, it stresses the  control-oriented nature  of the model proposed in Section~\ref{sec:CTMs_dyn}. In fact, the dynamics comply with the classical macroscopic traffic models and, at the same time, allow for a simple interconnection with control schemes.

\section{Simulations}
\label{sec:sims}
In this section, we analyse the \gls{CTMs} in the case of single and multiple service stations and, in particular, we study the effect that the model's parameters have on the overall traffic congestion. We consider a highway stretch divided in $N=9$ cells. The  parameters associated with the cells are reported in Table~\ref{tab:CTM_param}. The values has been identified from the data extracted from a stretch of the A$13$ highway in the Netherlands. To show and describe more clearly the effects of  the service stations on the traffic evolution, we assume that there are no on- and off-ramps. Given the parameters of the considered cells, the \gls{TTT} for the highway stretch in the case of free flow is $2.15$ min. The additional travel time during the time interval $k$ due to congestions can be computed as 
\begin{equation}
    \label{eq:Delta}
    \Delta(k) \coloneqq \sum_{i\in\mc N} \dfrac{L_i}{v_i(k)} - \dfrac{L_i}{\overline v_i},
\end{equation}
where $v_i(k)$ is the actual velocity in cell $i$ during $k$. In the following, we denote by $\Delta_0$ the quantity in \eqref{eq:Delta} obtained when there is no service station. During peak congestion we have $\max(\Delta_0(k))= 56$ s, which corresponds to an increment of the \gls{TTT} of $41.5\%$. 
The value of $\Delta $ is then a good indicator of the overall traffic congestion.  Notice that a reduction of  $\Delta$ implies, in turn, that the \gls{TTT} of the vehicles decreases.% since the highways operated closer to a free-flow condition. 

We perform the simulations over a time horizon of $3$h and consider a time interval of $T= 10$ s, so $k\in[0,1080]$. To better study the model's features, we consider a simplified piece-wise linear  flow entering the first cell that is defined for all $k$   as
$$ \phi_1(k)\coloneqq \max( 500, -7.04 |k-540|+2400).$$
We design $\phi_1$ to resemble the flow appearing during a typical morning rush hour in the considered highway stretch (a scaled version of $\phi_1$ is depicted in Figure~\ref{fig:e}). In this configuration, the period of high vehicles inflow lasts $1.5$h.
\begin{table}[t]
\caption{\gls{CTMs} highway cells' parameters}
\centering
\begin{tabular}{|c| c c c c c|}
\hline[2pt]
$i$ & $L_i$ & $\overline{v}_i$ & $w_i$ & $ q_i^{\max}$ & $\rho_i^{\max}$ \\ %table content
\hline
$1$ & 0.5 & 114 & 32.7 & 2511 &    97.1 \\
$2$ & 0.5 & 114 & 29.6 & 2472 &   105.7 \\
$3$ & 0.5 & 114 & 31.3 & 2338 & 95.1 \\
$4$ & 0.5 & 114 & 26.7 & 2310 & 106.7 \\
$5$ & 0.5 & 113 & 27.8 & 2337 & 104.8 \\
$6$ & 0.36 & 112 & 26.2 & 2343 & 110.2 \\
$7$ & 0.37 & 111 & 20 & 2136 & 126 \\
$8$ & 0.41 & 109 & 26.4 & 2317 & 108.9 \\
$9$ & 0.39 & 103 & 20.9 & 2111 & 
  121.6 \\
\hline
\end{tabular}
\label{tab:CTM_param}
\end{table}

\subsection{Single service station}
First, we assume the presence of a single service station between cells $2$ and $4$ and  explore the effects that it has on traffic congestion for different values of $\beta^\tup{s}$ and $\delta^\tup{s}$. As performance index we use $$\pi \coloneqq \dfrac{\max_k(\Delta_0(k)) - \max_k(\Delta(k))}{\max_k(\Delta_0(k))},$$ which denotes the percentage of peak congestion reduction. If $\pi$ value is $1$, then the introduction of the service stations completely eliminate the traffic congestion, while if it is $0$ there is no improvement compared to the case with no service station. Here, the priority of the main stream is set to \mbox{$p_4^\tup{ms}=0.97$.}

\begin{figure}[t]
\centering
\includegraphics[ width=\columnwidth]{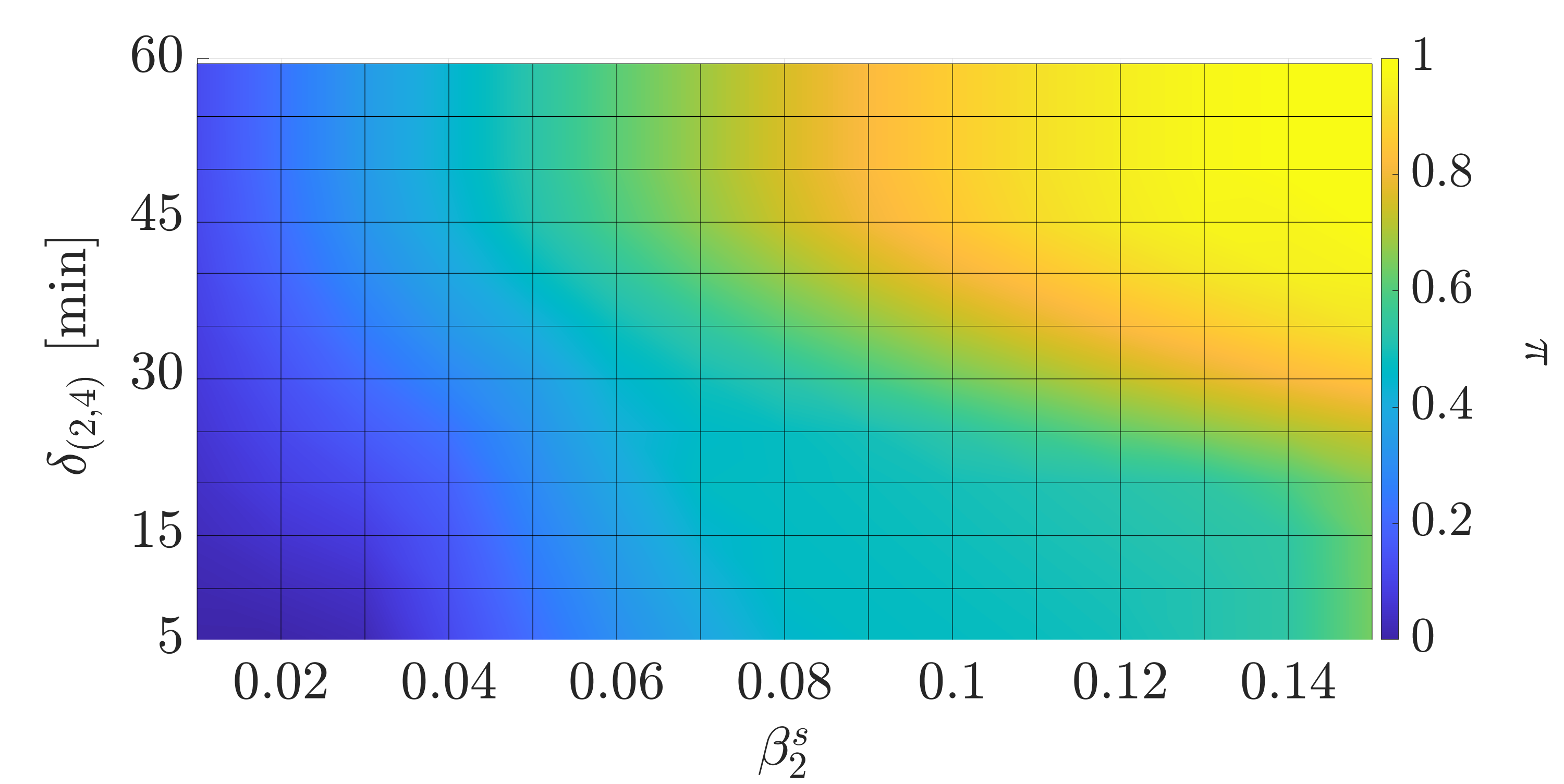}
\caption{The value of $\pi$ for a single service station and different values of $\delta_{(2,4)}\in[5,60]$ and \mbox{$
\beta_2^\tup{s}\in[0.01,0.15]$.}}\label{fig:delta_cmp}
\end{figure}

In Figure~\ref{fig:delta_cmp}, it can be noticed that an increment in $\beta^\tup{s}_2$ has a greater effect on the congestion than an increment  in $\delta$. In fact, even for $\delta_{(2,4)} = 5$ min, we achieve \mbox{$\pi = 0.64$} for $\beta^\tup{s}_2=0.15$ and  \mbox{$\pi = 0.30$} for $\beta^\tup{s}_2=0.06$. These correspond to $\max(\Delta(k))$ being equal to  $17$ s and $39$ s, respectively. As expected, the longer the drivers stop at the service station the higher the effect is on the traffic. In fact, if $\delta_{(2,4)} = 40$ min, then \mbox{$\pi = 0.97$} for $\beta^\tup{s}_2=0.15$ and  \mbox{$\pi = 0.54$} for $\beta^\tup{s}_2=0.06$.
\begin{figure}[t]
\centering
\includegraphics[ width=\columnwidth]{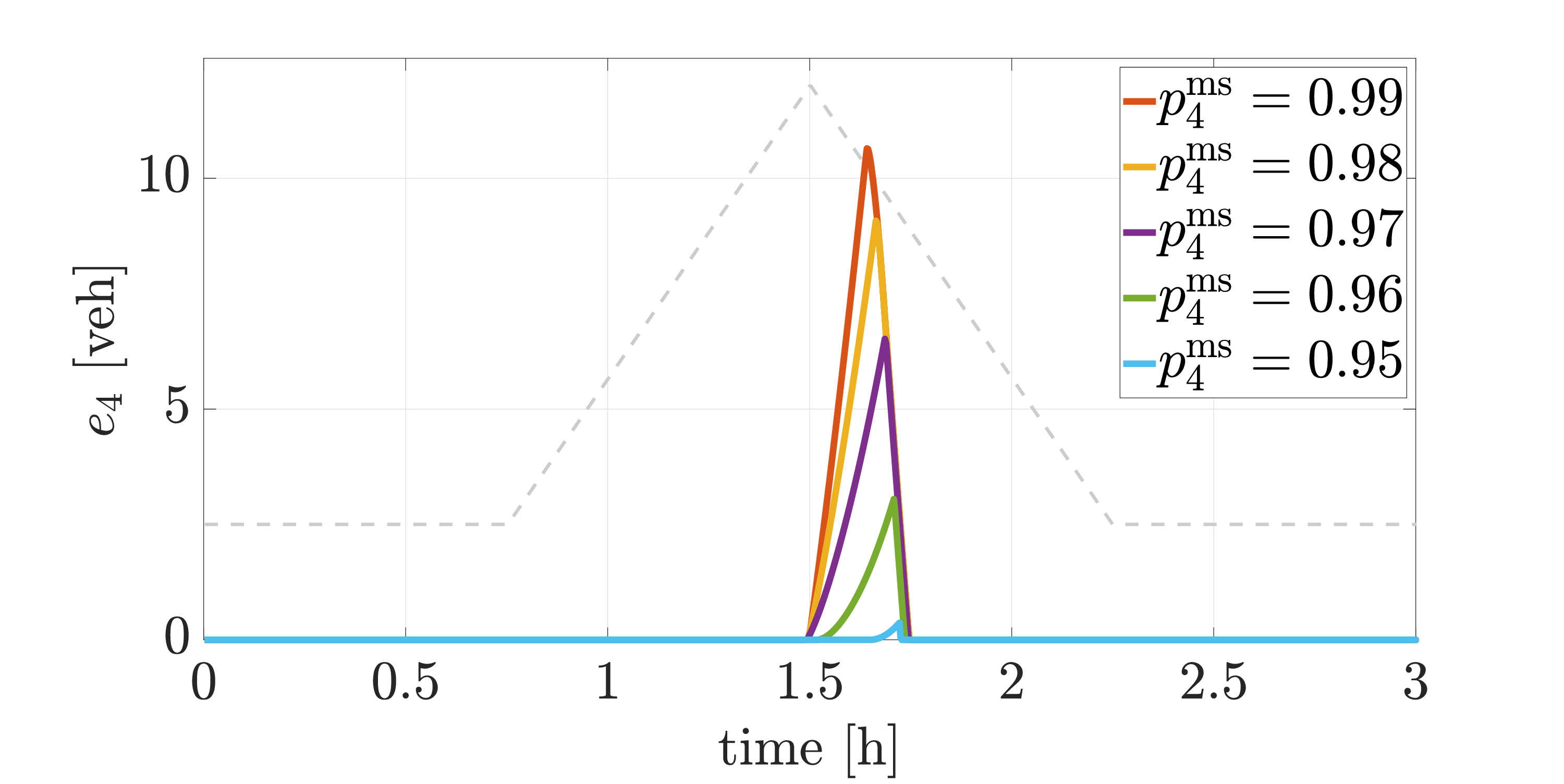}
\caption{The evolution of the queue of vehicles $e_4(k)$ waiting for exiting the service station for different values of priority of the main stream $p_4^\tup{ms}\in[0.95,0.99]$. The dashed gray line is a scaled version of $\phi_1(k)$.}\label{fig:e}
\end{figure}

Notice that a high  $\beta_2^\tup{s}$ might lead to an undesirable number of drivers waiting for merging back into the main stream. This issue can be amplified or mitigated by varying the priority $p_4$. To study this phenomena, we  examine evolution of $e_4(k)$ over time fro different values of $p_4^\tup{ms}$. 
We explore this scenario in Figure~\ref{fig:e}, where we plot the value of $e_4(k)$ in the case in which the priority $p_4$ varies from $0.95$ to $0.99$ and we choose $\delta_{(2,4)}=15$ min and $\beta_2^\tup{s}=0.05$. The maximum value of $e_4$ is always reached after $\delta_{(2,4)}$ from the peak of $\phi_1$. The maximum number of vehicles simultaneously  waiting  for merging is $e_4=11$ when $p_4=0.99$ while only $e_4=1$ if $p_4=0.95$. A higher priority usually leads to a reduction of   $\Delta$, since during congested periods the flow of vehicles entering the service station is bigger than the one exiting it.  This positive effect may be overshadowed by higher queues at the service stations that  discourage the single drivers from actually stopping and thus decreasing $\beta^\tup{s}$ and increasing $\Delta$. To define the  coupling among these two variables, one should model the decision-making process carried out by the drivers, as discussed in Section~\ref{sec:control}.    
%\smallskip
%\begin{remark}[Long-lasting congestion]\label{rem:long_congestion}

Traffic congestion throughout the days may  differ in nature and duration. This greatly affects the impact that a service station has on such a  traffic. In fact, if the period of maximum congestion is long and the main stream does not have a high priority, then the flow of vehicles exiting the service station may have a detrimental effect leading to an increment of the peak congestion. This emphasizes the necessity of a control action that coordinates the flows entering and exiting the service station during these more challenging scenarios. 
%\end{remark}

\begin{figure}[t]
\centering
\includegraphics[ width=\columnwidth]{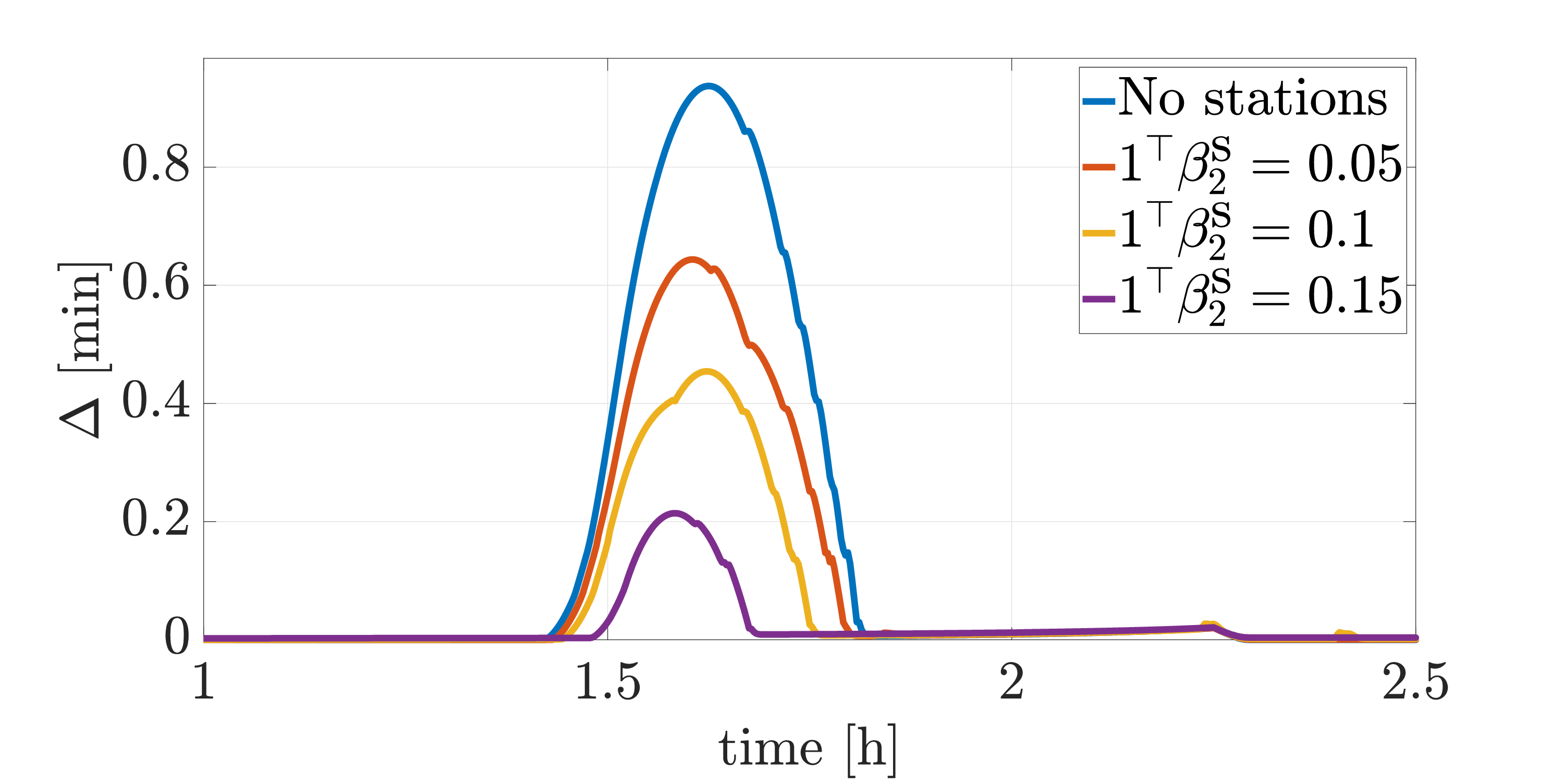}
\caption{The value of $\Delta(k)$ for different flows of vehicles entering the multi-purpose service station, i.e., $\beta_2^\tup{s}=[0.0225, 0.0225, 0.005]$ is plot in red, $[0.045, 0.045, 0.01]
$ in yellow, and $[0.0675, 0.0675, 0.015]$ in purple.}\label{fig:different_beta}
\end{figure}
\begin{figure}[t]
\centering
\includegraphics[ width=\columnwidth]{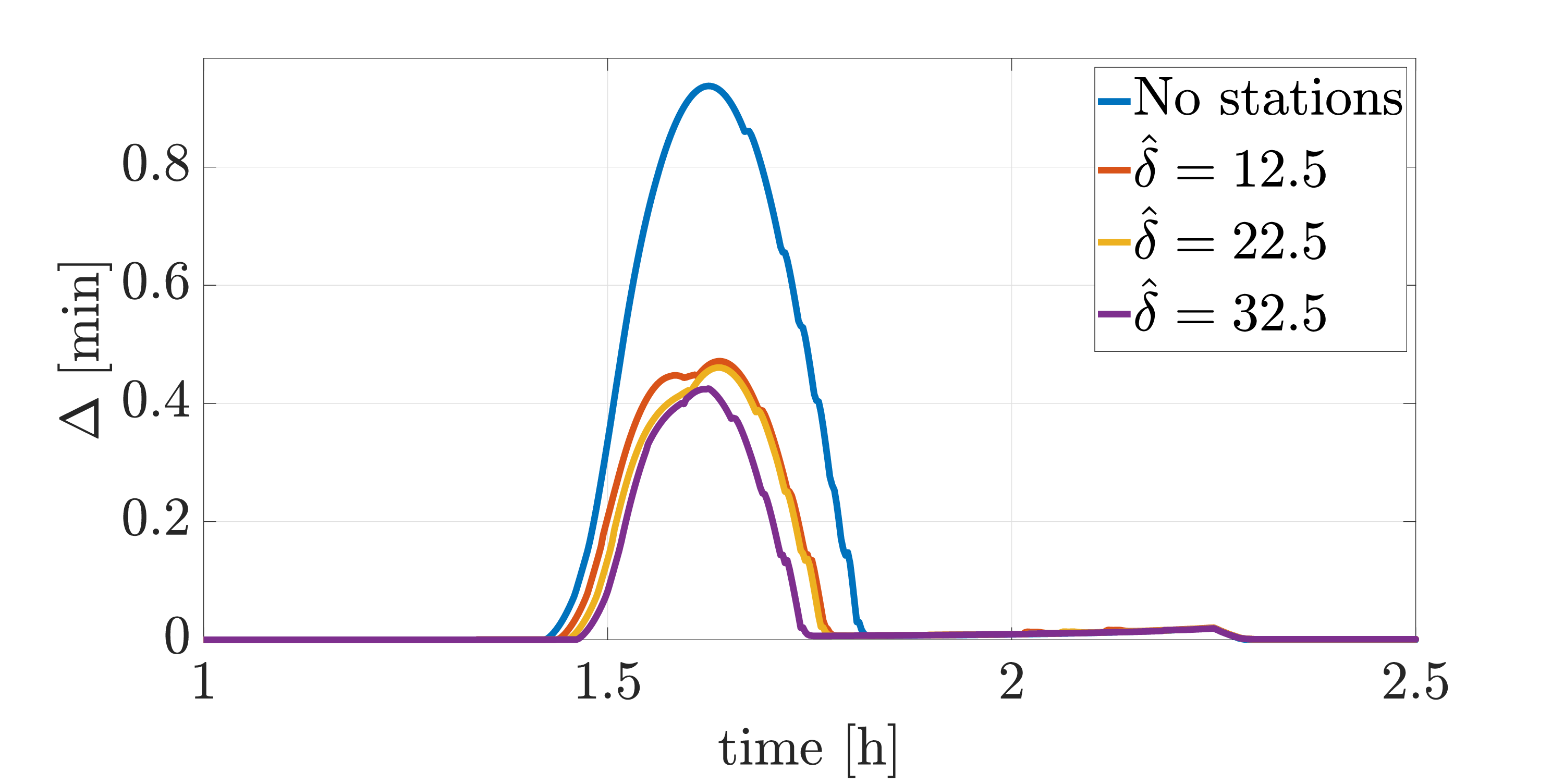}
\caption{The value of $\Delta(k)$ for different values of \mbox{$\hat\delta_{(2,4)}=(\beta_2^{\tup{s} \top} \delta_{(2,4)})/(\bs 1^\top \beta_2^{\tup{s}} )$}. Namely, $\delta_{(2,4)}  = [5, 15, 30] $ is plot in red, $[15, 25, 40]$ in yellow, and $[25, 35, 50]$ in purple.}\label{fig:different_delta}
\end{figure}
\subsection{Multiple-purpose service station}
Next, we consider the case of a multi-purpose service station, as depicted in Figure~\ref{fig:CTMs_conf}.a, placed  between cells $2$ and $4$. The drivers  can choose among three different services leading to distinct time periods spent at the station, so $\beta^\tup{s}\in\R^3$ and $\delta\in\R^3$. In the following, we  use $p_4=[0.97\,0.1\,0.1\,0.1]^\top$ where  $p_4^\tup{ms}=0.97$.

Firstly, we assume that the service station offers three different services which generally require the user to stop for an average period of $5$, $15$ and $30$ min, respectively. We assume that the service requiring more time is used less often than the other two, i.e., the associated $\beta^\tup{s}$ is smaller.    

In Figure~\ref{fig:different_beta}, we show the effect that this service station has on the traffic congestion in the case of different entering flows. We simulate three scenarios where  the total flow entering the service station $\bs 1^\top \beta^\tup{s}_2$ increases of  $5\%$ every time, going from $0.05$ to $0.15$. We obtain that $\pi=31.3$ for $\bs 1^\top \beta^\tup{s}_2=0.05$, $\pi=51.5$  for $\bs 1^\top \beta^\tup{s}_2=0.10$, and $\pi=77.1$ for $\bs 1^\top \beta^\tup{s}_2=0.15$. These values are similar to the ones that we would obtain in the case of a single service station with $\beta^\tup{s} = \bs 1^\top \beta^\tup{s}_2$. They show a steep reduction of the time spent in the traffic  congestion due to the presence of the service stations. 

In Figure~\ref{fig:different_delta}, we analyze the opposite scenario, that is a case in which the  inflow is fixed $\beta_2^\tup{s} = [0.035\,0.035\,0.01]^\top$ and the time spent at the service station increases.  We define the average time spent at the service station weighted by the percentage of vehicles using it as $\hat\delta_{(2,4)}=(\beta_2^{\tup{s} \top} \delta_{(2,4)})/(\bs 1^\top \beta_2^{\tup{s}} )$.  We consider three cases and the increment is of $10$ min for each one of the offered services, therefore $\hat\delta_{(2,4)}$ is $12.5$ min, $22.5$ min, and $32.5$ min, respectively. The improvements in the three cases is $\pi=0.49$, $\pi=0.51$, and $\pi=0.55$, respectively. Also in this case, the results are akin to those that can be obtained in the case of a single station with $\beta_2^\tup{s}=0.08$ and $\delta_{(2,4)} = \hat \delta_{(2,4)}$. It is remarkable that a reduction of almost $50\%$ of $\Delta$ is achieved with the smallest $\hat \delta_{(2,4)}=12.5$ min.

The results above align with the findings discussed in the case of single station. In fact, the effect of increasing the flow of vehicles entering the station has a greater effect on traffic congestion than the increment of the time spent in it.

\section{Conclusion and outlooks}
The \gls{CTMs} model is a novel macroscopic traffic model  based on the \gls{CTM}. It is particularly suited to describe  the traffic on highways (or simple routes) in which there are  service stations that can affect the dynamics. The flexibility of the model allows to easily describe several different scenarios like multi-modal service stations in which drivers stop for different reasons. Interestingly, the model shows that the introduction of a service station can reduce traffic congestion, in the case of a short traffic congestion, and it highlights that the number of vehicles stopping is more relevant than the time spent at the service station by the drivers.
The dynamics of the model have been developed to be easily interconnected with many  classical control schemes used in the literature to perform traffic management.

Being this the first paper introducing the model it favors several extensions for future research.
First, we want to include in the model the capacity-drop effect. It can  increase  \gls{CTMs} accuracy in describing how the merging back of the vehicles affect traffic conditions. It may also lead to significant differences between the case of a single and multiple stations. It is interesting to extensively study the  model via simulations to characterize the effect of different class of service stations in different traffic scenarios, and validate these findings by means of micro simulators such as Aimsun or SUMO.

\balance
\bibliographystyle{IEEEtran}
\bibliography{19_CDC_PEV_MC,library_CC}

\begin{thebibliography}{10}
\providecommand{\url}[1]{#1}
\csname url@samestyle\endcsname
\providecommand{\newblock}{\relax}
\providecommand{\bibinfo}[2]{#2}
\providecommand{\BIBentrySTDinterwordspacing}{\spaceskip=0pt\relax}
\providecommand{\BIBentryALTinterwordstretchfactor}{4}
\providecommand{\BIBentryALTinterwordspacing}{\spaceskip=\fontdimen2\font plus
\BIBentryALTinterwordstretchfactor\fontdimen3\font minus
  \fontdimen4\font\relax}
\providecommand{\BIBforeignlanguage}[2]{{%
\expandafter\ifx\csname l@#1\endcsname\relax
\typeout{** WARNING: IEEEtran.bst: No hyphenation pattern has been}%
\typeout{** loaded for the language `#1'. Using the pattern for}%
\typeout{** the default language instead.}%
\else
\language=\csname l@#1\endcsname
\fi
#2}}
\providecommand{\BIBdecl}{\relax}
\BIBdecl

\bibitem{eu:2019:traffic}
A.~Schroten, H.~van Essen, L.~van Wijngaarden, D.~Sutter, and E.~Andrew,
  ``Sustainable transport infrastructure charging and internalisation of
  transport externalities: Executive summary,'' European Commission, Tech.
  Rep., May 2019.

\bibitem{barth:2009:traffic}
M.~Barth and K.~Boriboonsomsin, ``Traffic congestion and greenhouse gases,''
  \emph{Access Magazine}, vol.~1, no.~35, pp. 2--9, 2009.

\bibitem{ganine:2017:resilience}
A.~A. Ganin, M.~Kitsak, D.~Marchese, J.~M. Keisler, T.~Seager, and I.~Linkov,
  ``Resilience and efficiency in transportation networks,'' \emph{Science
  Advances}, vol.~3, no.~12, 2017.

\bibitem{lighthill:1955:LWR}
M.~J. Lighthill and G.~B. Whitham, ``On kinematic waves ii. a theory of traffic
  flow on long crowded roads,'' \emph{Proceedings of the Royal Society of
  London. Series A. Mathematical and Physical Sciences}, vol. 229, no. 1178,
  pp. 317--345, 1955.

\bibitem{ferrara2018freeway}
A.~Ferrara, S.~Sacone, and S.~Siri, \emph{Freeway traffic modelling and
  control}.\hskip 1em plus 0.5em minus 0.4em\relax Springer, 2018.

\bibitem{daganzo:1995:CTM_part2}
C.~F. Daganzo, ``The cell transmission model, part ii: Network traffic,''
  \emph{Transportation Research Part B: Methodological}, vol.~29, no.~2, pp. 79
  -- 93, 1995.

\bibitem{Kontorinaki:2017:CTM_capacity_drop}
M.~Kontorinaki, A.~Spiliopoulou, C.~Roncoli, and M.~Papageorgiou, ``First-order
  traffic flow models incorporating capacity drop: Overview and real-data
  validation,'' \emph{Transportation Research Part B: Methodological}, vol.
  106, pp. 52--75, 2017.

\bibitem{gomes:2006:asymmetric_CTM}
G.~Gomes and R.~Horowitz, ``Optimal freeway ramp metering using the asymmetric
  cell transmission model,'' \emph{Transportation Research Part C: Emerging
  Technologies}, vol.~14, no.~4, pp. 244--262, 2006.

\bibitem{kamonthep:2010:multiclass_CTM}
K.~Tuerprasert and C.~Aswakul, ``Multiclass cell transmission model for
  heterogeneous mobility in general topology of road network,'' \emph{Journal
  of Intelligent Transportation Systems}, vol.~14, no.~2, pp. 68--82, 2010.

\bibitem{goodwin:2008:traffic_soft_measures}
S.~Cairns, L.~Sloman, C.~Newson, J.~Anable, A.~Kirkbride, and P.~Goodwin,
  ``Smarter choices: Assessing the potential to achieve traffic reduction using
  `soft measures','' \emph{Transport Reviews}, vol.~28, no.~5, pp. 593--618,
  2008.

\bibitem{cenedese:2019:PEV_MIG}
C.~{Cenedese}, F.~{Fabiani}, M.~{Cucuzzella}, J.~M.~A. {Scherpen}, M.~{Cao},
  and S.~{Grammatico}, ``Charging plug-in electric vehicles as a mixed-integer
  aggregative game,'' in \emph{58th IEEE Conference on Decision and Control},
  2019, pp. 4904--4909.

\bibitem{J_Hiskens_2013}
Z.~{Ma}, D.~S. {Callaway}, and I.~A. {Hiskens}, ``Decentralized charging
  control of large populations of plug-in electric vehicles,'' \emph{IEEE
  Transactions on Control Systems Technology}, vol.~21, no.~1, pp. 67--78,
  2013.

\bibitem{cenedese:arxiv:highway_part_I}
C.~Cenedese, M.~Cucuzzella, J.~M.~A. Scherpen, S.~Grammatico, and M.~Cao,
  ``{Highway Traffic Control via Smart e-Mobility -- Part I: Theory},''
  \emph{arXiv e-prints}, Feb. 2021.

\bibitem{cenedese:arxiv:highway_part_II}
C.~{Cenedese}, M.~{Cucuzzella}, J.~M.~A. {Scherpen}, S.~{Grammatico}, and
  M.~{Cao}, ``{Highway Traffic Control via Smart e-Mobility -- Part II: Dutch
  A13 Case Study},'' \emph{arXiv e-prints}, Feb. 2021.

\bibitem{SIRI2021109655}
S.~Siri, C.~Pasquale, S.~Sacone, and A.~Ferrara, ``Freeway traffic control: A
  survey,'' \emph{Automatica}, vol. 130, p. 109655, Aug. 2021.

\bibitem{depalma:2011:pricing}
A.~{de Palma} and R.~Lindsey, ``Traffic congestion pricing methodologies and
  technologies,'' \emph{Transportation Research Part C: Emerging Technologies},
  vol.~19, no.~6, pp. 1377--1399, 2011.

\bibitem{ferguson:2020:carrots_or_sticks}
B.~L. Ferguson, P.~N. Brown, and J.~R. Marden, ``Carrots or sticks? the
  effectiveness of subsidies and tolls in congestion games,'' 2020, pp.
  1853--1858.

\bibitem{Han:2017:CTM_resolving_freeway_jam}
Y.~Han, A.~Hegyi, Y.~Yuan, S.~Hoogendoorn, M.~Papageorgiou, and C.~Roncoli,
  ``Resolving freeway jam waves by discrete first-order model-based predictive
  control of variable speed limits,'' \emph{Transportation Research Part C:
  Emerging Technologies}, vol.~77, pp. 405--420, 2017.

\bibitem{Piacentini:2019:VACS_congestion_dissipation}
G.~Piacentini, M.~\v{C}i\v{c}i\v{c}, A.~Ferrara, and K.~Johansson, ``Vacs
  equipped vehicles for congestion dissipation in multi-class ctm framework,''
  \emph{2019 18th European Control Conference (ECC)}, pp. 2203--2208, 2019.

\bibitem{papageorgiou:1997:ALINEA}
M.~Papageorgiou, H.~Hadj-Salem, and F.~Middelham, ``Alinea local ramp metering:
  Summary of field results,'' \emph{Transportation Research Record}, vol. 1603,
  no.~1, pp. 90--98, 1997.

\bibitem{cenedese:2019:RBR}
A.~{Govaert}, C.~{Cenedese}, S.~{Grammatico}, and M.~{Cao}, ``Relative best
  response dynamics in finite and convex network games,'' in \emph{58th IEEE
  Conference on Decision and Control}, 2019, pp. 3134--3139.

\bibitem{cenedese:2021:EJC}
C.~Cenedese, G.~Belgioioso, S.~Grammatico, and M.~Cao, ``An asynchronous
  distributed and scalable generalized nash equilibrium seeking algorithm for
  strongly monotone games,'' \emph{European Journal of Control}, vol.~58, pp.
  143--151, 2021.

\bibitem{berridge:2000:reward}
K.~C. Berridge, ``Reward learning: Reinforcement, incentives, and
  expectations,'' in \emph{Psychology of learning and motivation}.\hskip 1em
  plus 0.5em minus 0.4em\relax Elsevier, 2000, vol.~40, pp. 223--278.

\bibitem{kreps:1997:extrinsic_incentives}
D.~M. Kreps, ``Intrinsic motivation and extrinsic incentives,'' \emph{The
  American Economic Review}, vol.~87, no.~2, pp. 359--364, 1997.

\bibitem{hess:2012:optimal_CS_pos}
A.~Hess, F.~Malandrino, M.~B. Reinhardt, C.~Casetti, K.~A. Hummel, and J.~M.
  Barcel\'{o}-Ordinas, ``Optimal deployment of charging stations for electric
  vehicular networks,'' in \emph{Proceedings of the First Workshop on Urban
  Networking}, ser. UrbaNe '12.\hskip 1em plus 0.5em minus 0.4em\relax
  Association for Computing Machinery, 2012, p. 1–6.

\end{thebibliography}

\end{document}